# CODE REVIEW AND COOPERATIVE PAIR PROGRAMMING BEST PRACTICE

Qiang Fu[1], Francis Grady[1], Bjoern Flemming Broberg[1], Andrew Roberts[1], Geir Gil Martens[1], Kjetil Vatland Johansen[1], Pieyre Le Loher[1]

[1] Schlumberger Information Solutions AS, Stavanger, Norway

Qfu3@slb.com

## ABSTRACT

*We need ways to improve the code quality. Programmers have different level of tenure and experience. Standard and programming languages change and we are forced to re-use legacy code with minimum revision. Programmers develop their habits and can be slow to incorporate new technologies to simplify the code or improve the performance.*

*We rolled out our customized code review and pair programming process to address these issues. The paper discusses the about the improvement of mandatory code review and pair programming practiced in the commercial software development, and also proposes effective approaches to customize the code review and pair programming to avoid the pitfalls and keep the benefits.*

## 1. INTRODUCTION

Several common issues can be noted from the programming practice of working on an existing big software: 1) outdated routines/patterns are still used, 2) some code does not follow the industrial standards, 3) changes are not properly documented. The importance of code quality can never be underestimated even for a deadline-driven world.

Code review, a manual inspection of source code by developers other than the author, is a common software engineering practice employed in industrial contexts and is recognized as a valuable tool for reducing defects and improving quality. The policy of 100 percent code review has been implemented/discussed in many commercial software projects.

Classical pair programming is an agile software development technique in which two programmers work together at one workstation [1]. Traditionally, one programmer writes code while the other reviews each line of code as it is typed in. The two programmers switch roles frequently. Some obvious benefits can be achieved with pair programming: 1) fewer bugs, 2) lower cost on production maintenance, and 3) knowledge transfer [2, 3]. Another benefit is that both developers acquire a good understanding of all the written code; they know what the design choices were and how the code works. From many aspects, this reduces the fragmentation of knowledge within a team.

Another agile software development technique, pair programming is also becoming increasingly popular in the software industry. It is more suitable for centrally-located team than the geographically-distributed team. It is commonly considered that pair programming can get more maintainable design with better quality, but in real working environment it often trapped in some pitfalls [4, 5]:

1) Discourages introversion. The coder must "program aloud" while the reviewer listens. Some developers will not raise concerns or suggest corner cases, thus turning the pair programming into "solitary programming" with automatic code review, which wastes resources.
2) Prevents creativity. Contrary to the value of "group brainstorming", creative work sometimes requires independence and autonomy. In pair programming, developers must be able to convince a partner of the merits of an idea. This requires talking through the implementation
3) Step by step and risking being judged if the idea fails.
4) Tiring practice. A good pair programming session is intense and mentally demanding. Programmers have reported significant exhaustion after just a few hours. This is a common observation, even from the most experienced practitioners and the advocates of pair programming.
5) Demanding balance maintenance. Pair programming can cost more work-hours than solitary programming to produce the same feature if the cooperation is not planned properly. A balance must be maintained carefully between the quality of code and the increased programming cost.

Mandatory code review and pair programming are being practiced in our team recently. Based on the actual circumstance of our team, the traditional code review and pair programming are tailored to get the advantages and avoid the pitfalls mentioned above.

## 2. CODE REVIEW

Mandatory code review was introduced in our team in July 2016. Although our main motivation for conducting code reviews was finding bugs, we found that reviews brought several additional benefits including knowledge transfer, increased team awareness and the creation of more elegant solutions.

From the outset, we established some principles:

1) Programmer reviewing. Code should be reviewed by active programmers, not the managers.
2) Rotating reviewer. Many code review guidelines recommend that the original author of a piece of code perform the review of any subsequent changes; in our case, that is largely impossible. Team and code ownership changes mean that the original author may work in a different team by the time the code is reviewed. Instead, we have introduced a simple rota for performing reviews. Every week, one developer is "on duty" for reviewing changes from all other developers.
3) Responsibility on reviewer. Reviewer takes the full responsibility for the changes.

To help improve review consistency, we have agreed on a checklist for both the reviewers to reference and programmers could use to recognize and resolve the issues in the code

 (Figure 1), and two reviewers are required when new team members join the team. This enables us to verify that key code goals such as readability, maintainability, and functionality are met.

| | | |
|---|---|---|
| 1 | Run the static code analysis tool (Resharper) before sending it to code review. Not necessary to follow each recommendation but follow which ever make | ☐ |
| 2 | 'null' check needs to be performed wherever applicable to avoid the Null Reference Exception at runtime. | ☐ |
| 3 | Use the extension methods, utility methods etc where ever possible. (For this you need to look into the existing classes) | ☐ |
| 4 | Create the resource to facilitate the Multilanguage feature. Add resource for all the culture supported and at appropriate layer. | ☐ |
| 5 | Check the existing code and try to follow the pattern. | ☐ |
| 6 | Run and make sure the existing test (Unit, Performance test) are passing. | ☐ |
| 7 | If new Class is added make sure its in new file and the name space is inline with folder structure. | ☐ |
| 8 | Formatting of the code. Send the well formatted code at least for the change in the code so that code quality is improved. | ☐ |

Figure 1 Customized code review checklist

Since one of the potential issues with code reviews is the lag time that they introduce into the development cycle, we added informal requirements that the size of the code to be reviewed be kept small and that reviews are completed in under 1 hour.

Another common issue is the "inherited" code. Since our software contains a huge legacy code base, we have agreed on 1) when creating a new program base on "inherited" code, programmer should be responsible for the existing code incorporated, 2) we recognize the time pressure to go through the legacy code and have a up-to-date version. Proper test should be in place for the legacy part of the code.

The overview of the code reviews can be set up in the Team Foundation Server (TFS) dashboard (Figure 2).

## 3. COOPERATIVE PAIR PROGRAMMING

If two is good, is three better? To push the code quality even further, we also performed cooperative pair programming. The project on which we tried was the creation of a new public API. The requirements and acceptance criteria were relatively clear, so the implementation, proper tests, and sample codes were the main work. Two developers worked on the project together, and both had adequate understanding on the work, which reduced the amount of discussion needed. Therefore, instead of having two people working on the same computer side by side all day and swapping roles frequently, we tailored our usage as follows:

1. As with classical pair programming, we sit together and agree on the API details such as the names, parameters, constants, etc.
2. After the API details are decided, the developers work at separate computers. One person works on the API implementation, and the other works on the tests for the designed API.

3. At the end of each day, regardless of whether the implementation or tests were finished, the developers swap roles. The person who was working on the API implementation reviews the test code and continues the test implementation, and vice-versa.
4. Steps 2 and 3 are then repeated until the work is complete.

By following this cooperative pair programming model, we gained several advantages:

1. We performed detailed and in-depth code reviews, which led to fewer bugs. Unlike common code reviews, we developed a stronger understanding of the code and the frequent communication that was required made it easier to find some of the more obscure bugs.
2. We observed a clear improvement in the quality of the code, including better readability and less unnecessary and unused code.
3. By switching the roles, API implementation code and its test code received a more thorough review.
4. We perceived increased knowledge sharing because it was necessary to understand the code thoroughly to continue the work. Because the code was fresh in the one developer's mind, it was easier to explain the intent to the other developer in the pair.
5. Both developers retained autonomy and the ability to exercise creativity. Both were free to try an approach before having to convince the other developer.
6. We obtained 100% code coverage on testing. Both developers spent the same amount of time writing the unit/acceptance tests as writing the API implementation.

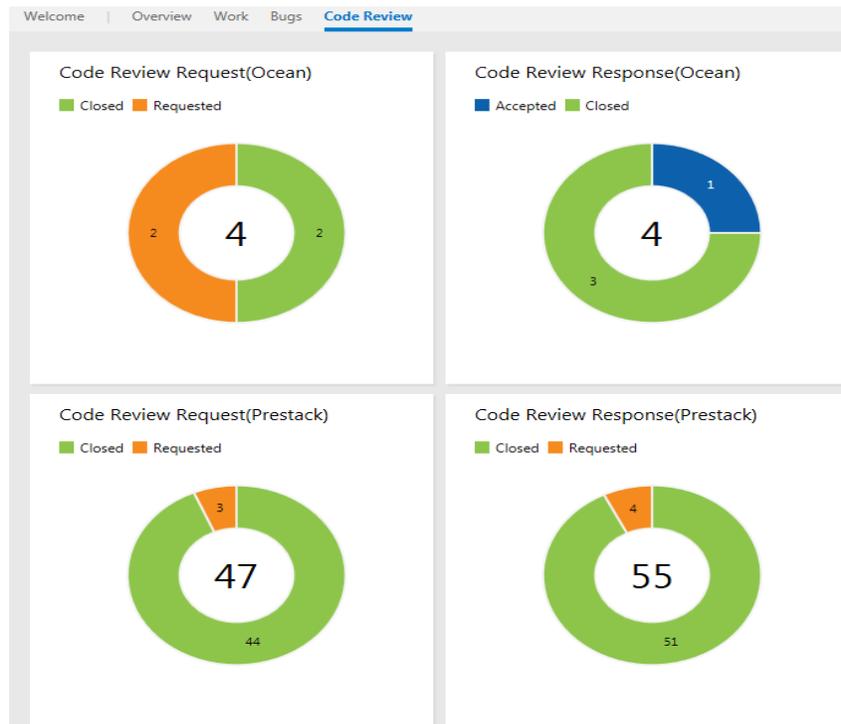

Figure 2 Code review in TFS dashboard

## 4. EXPECTATIONS AND OUTCOMES

After 4 months of mandatory code review, we have discovered that finding defects is not the only benefit of code review. Reinforced by a strong team culture around the reviews, we see several benefits:

**Productivity improvements:** We were concerned about the productivity in the beginning. But it actually improved the productivity. More defects discovered in the early stage, the fewer need to be tested later in the process. Of course, the review of the code takes time, but with modern tools like TFS the extra workload and level of disruption are kept to the minimal level.

**Code quality improvements:** A clear improvement on the code quality can be observed because of the mandatory review. The code review served as a reminder that other people will see and use the code and helped to uncover the "blind spots" when the programmers did not follow the coding standards. Improvements include better unit testing, fewer unnecessary changes and improved readability.

**Defect finding:** The detailed checklist and improved code quality enable us to discover obvious bugs such as exception handling, raw pointer misuser, typos and formatting mistakes. There was a gap between our expectations and reality in terms of the types of defects found. However, we still derive a benefit from catching the more obvious bugs earlier than in conventional programming.

**Knowledge transfer:** The team works on multiple separate projects. Code reviews help facilitate knowledge transfer between team members, not only helping to expose reviewers to a wider range of code, but also directing authors to other resources for learning how to solve some problems. In at least one case the process led to a mentor relationship between programmer and the reviewer who was helpful after the code review was done.

**Team awareness and transparency:** By performing mandatory code reviews, we not only keep the team generally aware of changes in the code, we also prevent anyone from adding low quality "Band-Aid" fixes to the code in secret.

From our cooperative pair programming experiment, we have discovered some conditions that effect the success of pair programming:

1) The maturity of the design
2) The comparative skill levels of the developers involved
3) The scale of the work, with the best scale being a task totalling at least two person-months estimated work.

## 5. RECOMMENDATIONS

The roll-out of code review and pair programming is promising. The feedback from the programmers and reviewers are generally positive. From the experience, we can offer several observations and recommendations:

**Customized checklist:** Each team should have tailored checklist according to its programming environment and team culture, and this checklist should be updated as the team and its projects change.

**Quality assurance:** Code reviews rarely result in identifying subtle bugs, so standard QA practices such as automated unit testing and acceptance tests should be maintained.

**Beyond defects:** Code reviews provide benefits beyond finding defects. They can be used to help standardize style, find alternative solutions and increase learning. These goals should guide code review policies.

**Customized pair programming:** Cooperative pair programming is just one of many possible customizations of pair programming. Depending on the circumstances, different variants of pair programming could be tried to provide an optimal balance between quality and cost.

**Pair rotation:** In relatively big engineering team, pair rotation could be considered rather than having assigned pairing partners working all the time. It can aid in introducing and training new team members.

**Authors**

**Qiang Fu** was born in China in 1977. He received the Ph.D degree from Imperial College London in 2010. He joined Schlumberger Information Solution AS in 2011 as senior software developer in Petrel Geophysics team. His main areas of research interest are software processing, developing, geophysics and geology.

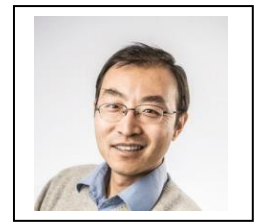

**Francis Grady** received his Master's degree in Computer Science from the University of Oxford in 2006. Since then he has been with Schlumberger, where he is currently a Senior Software Engineer. His interests include machine learning, high performance computing and code quality.

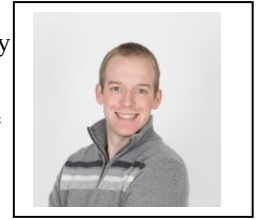

**Bjoern Flemming Broberg** joined Schlumberger in 2013 working as a Senior Software Engineer developing software. He has a master in industrial mathematics from Trondheim in Norway, and has more than 20 years of experience as an IT professional working as business analyst, IT architect, developer and IT project manager.

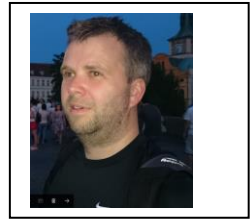

**Andrew Roberts** has worked for six years at Sclumberger as a Software Engineer, in development, build and configuration management, and testing roles. Prior to Sclumberger he was Software Consultant for over a decade in the mobile devices market working with such companies as

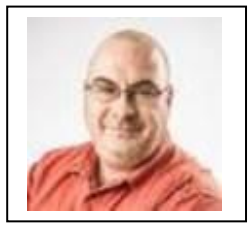

Motorola, Nokia, Panasonic, etc.

**Geir Gil Martens** was born in Bergen, Norway, 1960. After acquiring an undergraduate degree in computer science at Rogaland Distriktshøgskule, Norway. He joined Geophysical Company of Norway – GECO AS in 1985 to develop the Charisma II Seismic Interpretation Station. Over the years he have been involved with most aspects of software development and a multitude of more or less formalized development processes. He is currently working at Schlumberger SNTC as a senior software engineer on the Petrel system.

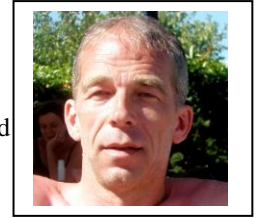

**Kjetil Vatland Johansen** has a M.Sc. degree in Technical Cybernetics from Norwegian University of Science and Technology. He has combined background from cybernetics with a passion for software development throughout the professional career. He was a developer in an C++/.Net environment for 15 years and then moved to project management.

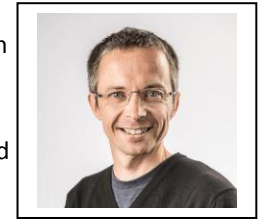